# Smart Jamming Attacks in 5G New Radio: A Review

Youness Arjoune
School of Electrical Engineering and Computer Science
University of North Dakota
Grand Forks, ND, USA
youness.arjoune@und.edu

Saleh Faruque
School of Electrical Engineering and Computer Science
University of North Dakota
Grand Forks, ND, USA saleh.faruque@und.edu

*Abstract*—The fifth generation of wireless cellular networks (5G) is expected to be the infrastructure for emergency services, natural disasters rescue, public safety, and military communications. 5G, as any previous wireless cellular network, is vulnerable to jamming attacks, which create deliberate interference to hinder the communication of legitimate users. Therefore, jamming 5G networks can be a real threat to public safety. Thus, there is a strong need to investigate to what extent these networks are vulnerable to jamming attacks. For this investigation, we consider the 3GPP standard released in 2017, which is widely accepted as the primary reference for the deployment of these networks. First, we describe the key elements of 5G New Radio (NR) architecture, such as different channels and signals exchanged between the base station and user equipment. Second, we conduct an in-depth review of the jamming attack models and we assess the 5G NR vulnerabilities to these jamming attacks. Then, we present the state-of-the-art detection and mitigation techniques, and we discuss their suitability to defeat smart jammers in 5G wireless networks. Finally, we provide some recommendations and future research directions at the end of this paper.

## I. INTRODUCTION

The fifth generation of wireless cellular networks, 5G, promises faster data rates and reliable service delivery. It is expected to enable many cutting-edge technologies such as internet-of-things (IoTs), self-driving cars, and smart cities. In 2017, 3GPP released the specification of 5G New Radio (NR), which has been the primary reference for the deployment of these networks. 5G NR architecture is built upon five fundamental pillars: New radio spectrum, massive MIMO/beamforming, multi-connectivity, network flexibility, and high level of security. 5G is operable on a new radio spectrum from below 1 GHz to up to 100 GHz. The 5G NR physical layer uses orthogonal frequency division modulation (OFDM) with a cyclic prefix on the downlink and either the OFDM or discrete Fourier transform-spread OFDM for uplink. The 5G NR frame is of 10 ms duration, in which there are ten sub-frames and fourteen OFDM symbols. 5G NR supports both the frequency division multiplexing FDD and time division multiplexing TDD modes [1].

As any wireless cellular networks, 5G networks are built upon open sharing in which the communication medium is the free space making them prone to interference, which is one of the fundamental causes of degradation of the performance of wireless networks. If the level of obstruction is high, the receivers are not able to decode the transmitted signals. This weakness can be used by some adversary nodes to cause intentional interference and hinder legitimate user's communication over specific wireless channels. This is well-known as jamming attacks.

Jamming attacks pose serious risks to public communication services [2], [3]. In early 1900, jamming attacks were used in military battles. Nowadays, jamming attacks can be launched to hinder public communication services. Several jammer devices are available in the market at a low cost. In addition, the most sophisticated jamming attacks can be implemented with a price as low as $1k\$$ using low-cost software-defined radio tools, and some primary programming skills. Furthermore, 5G is expected to be the infrastructure for emergency services, natural disasters rescue, public safety, and military communications making jamming attacks a real threat.

Therefore, one of the central requirements of 5G NR is a high level of security and resilience to jamming attacks [3]. 5G is expected to enhance the security of wireless networks and fix the breaches of long-term evolution (LTE) or 4G networks, especially the resilience to jamming attacks. In 2017, 3GPP released the 5G NR standard. Before deploying these cellular networks, one needs to investigate to what extent the 5G standard released by 3GPP is resilient to jamming attacks, and build security protocol that can be incorporated in this standard. Thus, it is essential to determine under which conditions (e.g., the power of the jammer, the duty cycle, etc.), the jammer can take down the communication channel; and to determine which anti-jamming techniques are suitable for 5G NR.

In this paper, we aim to provide an in-depth study of these issues, assess the risks of jamming attacks on 5G NR, and suggest some possible future research directions on how to efficiently tackle this problem. The main inputs of this article can be summarized as follows:

- Description of the key elements of 5G NR architecture
- Review of jamming attack models and strategies

- Assessment of jamming vulnerabilities of the 5G NR
- Review of detection and mitigation techniques and discussion of their suitability to defeat smart jammers in 5G
- Conclusions, recommendations, and future research directions

The remaining sections of this paper are arranged as follows. Section II describes the 5G NR architecture. Section III defines jamming attack strategies, types, and models. The overall 5G NR vulnerability to jamming attacks assessment is provided in Section IV. Section V revisits briefly the state-of-the-art anti-jamming and mitigation techniques and discusses their suitability for 5G NR. Section VI summarizes the results of this work and draws the conclusions.

## II. 5G NR Architecture

5G NR is operable on from low to very high-frequency bands (0.6-30GHz). It gives ultra-wide carrier bandwidth, which can be up to 100 MHz in below 6 GHz and up to 400 MHz in higher than 6 GHz. In 5G NR, there are several physical channels. For instance, for the downlink, there is downlink shared channel (PDSCH), Broadcast channel (PBCH), and downlink control channel (PDCCH). For the uplink, there is uplink shared channel (PUSCH), uplink control channel (PUCCH), and Random access channel (PRACH).

The physical layer of 5G NR includes many types of signaling reference signals and synchronization pilots exchanged on both downlink and uplink. For instance, the base station uses the primary synchronization signal (PSS) and secondary synchronization signal (SSS) for downlink frame synchronization and conveying cell-ID to user equipment (UE). The PSS has three possible combinations, while the SSS has 336 combinations. Each of the PSS and SSS consists of an m sequence of length 127, and is mapped to a set of 127 subcarriers within the same OFDM symbol, different OFDM symbols, respectively. The use of the Gold sequence, which is formed by combining two orthogonal m-sequences, enables the UE to differentiate between several base stations on the same carrier at a low signal-to-noise ratio (SNR) [4].

5G NR enables scalable NR numerology to address different radio spectrum, bandwidths, and services. For instance, subcarrier spacing (SCS) of 15, 30, 60, and 120 kHz is specified for macro coverage, small cell, indoor, and mmWave, respectively. 5G NR frame is similar to the one of 4G/LTE with some slight modifications. One slot in the 5G NR frame is composed of 14 symbols, and the slot length is dependent upon the CSC. Mini-slot is comprised of 2, 4, and 7 symbols, which can be allocated for shorter transmissions. Slots can be aggregated for more extended periods of communication. The OFDM symbol contains PSS, PBCH, and SSS.

For the coding schemes, 5G NR uses low-density parity check (LDPC) for the data channel and polar coding for the control channel. It has been shown that LDPC codes perform well when used for error correction for small chunks of data. Polar coding, on the other hand, can achieve performance close to the Shannon limit, but it has to be used with large pieces of data.

Another feature of 5G NR is the use massive Multiple input multiple output (MIMO) to enhance the coverage and the capacity of wireless cells [5]–[8].

## III. Jamming attacks

Jammers can be defined as malicious wireless nodes planted by an adversary to cause intentional interference in wireless cellular networks. Depending on its attacking strategy, one can identify several types of jammers. In the following, we provide the most common types of jammers and describe their strategies.

*1) Regular jammer:* In this type, the jammers tend to not follow any MAC protocol before continually injecting radio frequency signals without gaps in between, which can be either legitimate bit sequences or random bit sequences to interfere with legitimate transmitted signals over a wireless channel. Subsequently, these bits occupy the transmission channel to starve transmissions initiated by legitimate nodes. This type of attack uses enormous power, which drains the battery life of the malicious node due to its continuous transmission of radio signals. Regular jammers, consequently, require a high amount of power to carry out this attack. On the other hand, regular jammers do not need to monitor the activity of legitimate users [9]–[11].

*2) Delusive jammer:* In this type, known also as deceptive, jammers continuously inject legitimate sequence of bits into the communication channel. This type of jammer often misleads the receiver to believe that this is a message from a legitimate source. It forces the receiver to wait in the listening states. In comparison with regular jammer, delusive jammer tends to be quite challenging to detect because of the similarity between the fake signal and the legitimate one [10], [11].

*3) Random jammer:* Different from both regular and deceptive jammers, random jammers conserve their energy by alternating between active and idle states. During the jamming process, the malicious node jams for a predetermined period before turning off its radio. After a while, it reactivates the jamming process from the sleep mode and continually follows that pattern. During the jamming mode, it can exhibit either regular or deceptive jamming feature, while during the idle state, it conserves energy and therefore reducing its power consumption [9]–[11].

*4) Responsive jammer:* All of the three previous jamming strategies discussed before are active jammers, as they attempt to block the communication channel, regardless of the activity pattern of the legitimate nodes. An alternative to active jammers to reduce its power consumption is to be a quick responsive jammer, known also as reactive jammers, which can be a more power-efficient method. Responsive jammers continually monitor the communication channel, and transmit only of the transmitter is active [9], [10], [12], [13]. Responsive jammers minimize power consumption despite the monitoring activity for the power required is far less than the one necessary to jam a communication channel. For instance, the authors of [7] launched a jamming attacks using deep learning.

   5)   *Go-next jammer:* This jammer is selective because it targets one frequency channel at a time. If the transmitter detects the existence of a jammer over the frequency channel and hops to the next frequency, this kind of jammer follows on the transmitter and goes to the next frequency channel. Due to its selective nature, go-next jammer conserves its energy. Notwithstanding, if the transmitter performs fast rate frequency hopping, the jammer's energy can be wasted because of the successive hops [11].

   6)   *Control channel jammers:* This jammer targets the control channel to block the exchange between the transmitter and the receiver before initiation of the communication. Control channel jammer can be of several types and can cause a denial of service and even denies nodes access to the network [11], [14].

IV. Vulnerableness of NR to Jamming Attacks

*A. Vulnerableness of the PBCH to Jamming*

Base station PBCH are assigned symbols within two slots of each other if the carrier is below 3 GHz and within four slots if the carrier frequency is above 3 GHz [4]. As higher the sub-carrier spacing (SCS), the duration of one slot is smaller, and the selective jamming duty cycle is lower. Consequently, a selective jammer can target the PBCH using a shallow duty cycle as the symbols are close to each other in both cases. This design is a vulnerability in design even that the use of higher frequency does not propagate for a long-distance making the jammer getting closer to the mobile station to launch its attacks. A localization-based detection technique can identify the source of the jammer and stop it. However, if the jammer is mobile, the anti-jamming has to monitor the mobility of the jammer to detect the next jammer positions. The longer the monitoring process, the harmful the jamming is going to be.

*B. Vulnerableness of PDCCH to Jamming*

CORESET is a set of physical resources (i.e, a specific area on 5G NR Downlink Resource Grid) and a set of parameters that are used to carry PDCCH. It is equivalent to LTE PDCCH area (the first fourth OFDM symbols in a subframe). But in LTE PDCCH region, the PDCCH always spread across the whole channel bandwidth, but NR CORESET region is localized to a specific region in the frequency domain. Jamming PDCCH channel is far more complicated than jamming PBCH channel. To jam PDCCH, the jammer has to cram all the possible locations in which the PDCCH resides, assuming that the jammer does not have any knowledge of the CORESET freq-domain. However, the jammer can intercept and decode the CORESET freq-domain, which gives the jammer and advantage to jam specific sub-carrier, using a small duty cycle depending upon the value of CORESET-time-duration. The question is how long it can take to intercept and decode the CORESET.

*C. Vulnerableness of PUCCH to Jamming*

The PUCCH has an option for intra-slot hopping, which can provide some protection against selective jammer, but the robustness of this defense mechanism is dependent on the hopping rate. Also, this information is available to the jammer as the 5G standard is public. Thus, knowing the intra-slot hopping gives the jammer an advantage to jam PUCCH at low cost. Furthermore, the PUCCH is modulated with MPSK (m=2 or 4) and polar code or just repetition code as an error coding scheme depending on the number of bits to be transmitted. Polar codes are well known by their low protection against jamming attacks.

*D. Vulnerableness of RACH to Jamming*

The random access (RA) procedure is the uplink transmission of a random access preamble by the UE on a dedicated RACH. After the reception of a preamble, the base station estimates temporal synchronization parameters and allocates radio resources for further communication with the UE. The synchronization parameters and allocation of radio resources are then communicated to the UE that initiated the RA procedure within a specified time after the RA preamble transmission [15]. This information is broadcasted on PRACH, which takes the form of a Zadoff-Chu sequence that embeds a value used to identify the UE temporarily. Despite the large number of possible locations, and the high complexity needed to determine the positions in real-time, jamming PRACH is still feasible [4]. Furthermore, if the jammer does not succeed in determining these locations, it can flood the channel with an invalid preamble as the 5G NR does not specify what it should be done in this scenario.

*E. Vulnerableness of Massive MIMO to Jamming*

Many research studies (industrial and academic) showed that massive MIMO are vulnerable to jamming attacks. Jamming MIMO systems targets the channel estimation of these systems. By targeting the channel estimation procedure, an adversary may launch active jamming attacks against unsuspecting users. The authors of [16] presented several jamming methodologies for SVD-based MIMO systems, including a powerful and practical channel rank attack. The authors presented several attack strategies to undermine Alamouti STBCbased MIMO scheme. Such attacks have been proven feasible by way of analysis, simulations, and real-world experimentation. Additionally, the attack strategies presented are general and remains valid for massive MIMO systems. Therefore, preserving accurate channel estimation under jamming attacks is quintessential to gain the desired performance enabled by massive MIMO. Hence, there is an urgent need for developing techniques for accurate channel state estimation whose performance is not impacted by the presence of jammers, or at least it considers the presence of the jammer and estimate the channel state information from affected pilot samples.

*F. Robustness of 5G NR Channel Coding to Jamming*

3GPP specifications for 5G radio standard includes polar coding and LDPC coding techniques. Polar coding which uses the channel polarization to split the channel into good channel and worse channel and transmit only on the good channel, presents several advantages but have some limitations. LDPC codes on the other hand if used with large block, the complexity of the decoder increase exponentially. Most of the control channels use polar coding as error coding scheme, and the data channels use LDPC coding. It has been shown that polar coding is vulnerable to jammers. At $0 dB$ SNR, the bit error rate is so high. Likewise, LDPC coding are vulnerable to jamming attacks.

V. ANTI-JAMMING IN 5G

In this section, we review the anti-jamming techniques. We divided this section into three parts: the first part presents jamming detection methods, the second part deals with mitigation methods, and the last part provide a discussion on the effectiveness of these techniques in tackling jamming attacks in 5G.

*A. Detection of Jamming Attacks*

Detection of smart-jamming attacks is feasible by monitoring any excess amount of energy on a specific physical channel (e.g. using masking) or any sudden change in the performance of the communication over this channel. A common strategy in jamming detection is the use of a threshold with some performance metrics such as the packet delivery ratio (PDR), packet drop ratio (PDR), bit error rate (BER), and signal-to-noise ratio (SNR). These techniques monitor the level of these metrics during the absence and the presence of jamming attacks and set manually the threshold for detection. Threshold based detection are only efficient when we are dealing with constant jammer. In addition, because of the wireless environment dynamics, these methods have a high false alarm.

Another detection category is statistical based [17], [18]. The concept of these techniques often uses historical data and compute some statistic to distinguish jammed signal from a non-jammed signal. Statistical detection of jamming is investigated with different forms of jamming attacks, and can achieve high accuracy when dealing with constant jammers.

The last category is machine learning based. Several machine learning techniques such as random forest, decision tree, adaptive boosting, support vector machine, and expectation maximization are investigated in detecting jamming attacks [19]. Recently, deep learning which is a special case of machine learning is heavily investigated to detect jammers [7]. Deep learning can detect jammers with high accuracy. Nevertheless, deep learning presents some limitations. There is no public dataset that can be used to train machine learning models. Most of the proposed methods generated dataset using simulations and only a few papers have conducted real-world setup to collect data. It is hard to foretell the performance of these detection techniques under a real jamming attacks.

*B. Mitigation of Jamming Attacks*

5G networks are going to use frequency higher than 30 GHz. Jamming these bands are not likely to happen because the jammer need high level of power to jam these bands [4]. In addition, 5G is going to implement techniques such as direct sequence spread spectrum (DSSS) and frequency hopping spread spectrum (FHSS) [20]. One need to investigate to which extent these techniques can act against jamming at the 5G physical layer.

*1) Direct Sequence Spread Spectrum:* spread spectrum can provide protection against interfering jamming signals with finite power. This technique purposely makes the information bearing signal occupy a bandwidth larger that of the minimum necessary to transmit it. Thus, the signal is transmitted through the channel undetected by an eavesdropper. In direct sequence spread spectrum (DSSS), the data signal is multiplied with a pseudo-noise (PN) sequence. The data signal is a narrow band and the PN sequence is wideband making the product nearly have a spectrum as same as PN which plays the spreading code role. The resilience of this technique against jamming attacks depends upon the spreading factor. An example of the protection that DSSS provides is shown by the authors of [21], in which a BPSK modulated signal is considered and if $N = 4095$ and the BER is not to exceed $10^5$, the authors showed the data at the receiver can be detected reliably even the jamming power is more than 400 times the received signal power. This example shows that direct sequence spread spectrum is powerful against interference jamming. Nevertheless, DSSS has some limitations such as the larger bandwidth required and the complexity of this techniques which can be an obstacle to implement them in some wireless devices. In addition, it has been shown that time division multiple access which uses DSSS can be cracked on real-time and detect the PN code. by having access to such information, the jammer can launch a follow-on jamming to hinder the communication of its target as low cost.

*2) Frequency-hop Spread Spectrum:* DSSS techniques are powerful yet they impose some practical limits because of the capabilities of the physical devices used to generate the PN sequence. Specifically, it may turn that the generated processing gain is still not large enough to overcome the effects of some jammers, which is same cases, resort to different strategies. One way to get around this problem is by randomly hopping the data modulated carrier from one frequency to another [22], [23]. In this type of spread spectrum, the spectrum of the transmitted signal is spread sequentially instead of instantaneously. One characteristic of Frequency-hop is the hopping rate, based on which one can distinguish between two types:

- Slow-rate hopping: Several modulated symbols are conveyed within one frequency hop before it hops to the next frequency.
- Fast-rate hopping: is the converse of the slow-frequency hopping, which means one symbol rate is transmitted during several frequency hops.

Using frequency hopping has many limitations. For instance, using a slow-rate hopping does not provide a robust protection against smart jammer as this jammer can find the next hop before the transmitter switches to the next frequency; and using a fast-rate hopping can decrease the performance of the communication channel as it becomes hard to synchronize the transmitter and the receiver. In addition, frequency hopping requires a pre-shared key between the transmitter and the receiver to agree on the hopping pattern, exchanging the keys can be intercepted by an eavesdropper. The authors of [24] proposed a secretive adaptive frequency hopping scheme for 5G. The authors of [25] proposed a pseudo-random time hopping for anti-jamming in 5G wireless networks. The authors analytically evaluated the performance of the proposed scheme by determining the jamming probability, the switching rate, and bit error rate. The authors of [26] proposed a frequency hopping for 5G mmWave. Yet, there is need to evaluate the impact of frequency hopping on outage probability for 5G mmWave.

*3) Game Theory:* Game theory is another anti-jamming techniques that aim to find the optimal strategy to defeat jammers [8], [27]–[29]. Legitimate users can avoid the jamming attacks by proactively hopping among accessible channels and thereby minimizing the payoff function. The anti-jamming in this context is expressed as a game between the legitimate user and the jammer. Game theory can be used to find the optimal strategy to cope with a jammer such as hopping to the next frequency. Several researchers have shown that it is possible to achieve the Nash equilibrium, meaning that the transmitter can find the optimal strategy to cope with the jammer.

*4) Timing channels:* The timing channel restores the communication between legitimate users under jamming attacks instead of frequency hopping. The timing channel is reinstated over the jammed channel using the timing patterns of attacker [30], [30]. This information enables the transmitter to transmit only when the jammer is in the idle-state. The timing channel requires the detection step before the creation of the timing channel.

*5) UAVs and Reinforcement learning:* An unmanned aerial vehicle (UAV) aided 5G wireless communication framework is yet another anti-jamming strategy. The UAVs are used as a relay scheme if the base station is heavily jammed. The UAVs use deep reinforcement learning techniques to determine the optimal relay policy for mobile users in 5G cellular networks. Examples of solutions based on UAVs and Deep reinforcement learning have been proposed in [6], [31]–[34]. This solution could be useful because of the flexibility the UAVs give to the network to avoid the jammer. Nevertheless, this framework faces many challenges, and UAVs themselves are also vulnerable to jammers, and its power supply is limited.

*6) Suppression of Jammers:* Massive MIMO suppression is a potential technique that can be enhanced and used to deal with jamming, and it does not require any change in the 5G NR specification [35]. The solution to interference is to build robust channel coding schemes that can correct packets corrupted by the jammers. This strategy can exhaust the jammer.

To improve the resilience of 5G systems to jamming attacks, the authors of [36] proposed a jamming-resistant receiver scheme. The prominent feature of this proposed scheme is that, in the pilot phase, the base station estimates estimate both the jamming channel. The jamming channel estimate is then exploited to build linear receiver filters that reject the impact of the jamming signal. The authors of [37] proposed a mitigation technique based on random matrix theory.

*7) Scheduling and Deep Learning:* 5G wireless network functions SDN and NFV alongside deep learning can help in building intelligent dynamic radio resource allocation and scheduling, which can significantly reduce the risk of jamming attacks. For instance, the authors of [38] proposed the joint power control and scheduling problem in jammed networks under minimum QoS constraints without any prior about the jammer positions. If combined with Deep learning to learn the strategy of the jammer, scheduling can achieve better performance.

## C. Effectiveness of Mitigation Techniques and Future Research Directions

Direct spread spectrum techniques can achieve high protection against jamming attacks. However, the complexity of this technique can be an obstacle to the implementation of this technique. Frequency hopping techniques are not suitable to counter jammers because the latter can predict the next channels, and any exchange between the receiver and the transmitter can be intercepted [20]. Machine learning based anti-jamming schemes are not practical for some applications in 5G wireless networks as it requires long training time and it requires building large comprehensive dataset to have reliable detection accuracy. Timing channels can be reliable if combined with excellent detection technique. UAVs based mitigation is a promising solution, but further investigation of the practical issues should be considered.

Thus, further research studies on anti-jamming techniques are highly needed. The cyber-security requirement of 5G NR has to be embedded in the initial design of these networks. In this way, one can ensure a low-cost deployment, in contrast, to develop solutions to deal with future failures. For instance, base station must implement anti-jamming techniques. For example, if the exchange between the base station and user equipment, the base station should provide a spatial retreat, movement, time, and network reconfiguration.

Another future research direction is the use of a deep learning based approach can be used as anti-jamming. To train deep learning, a large comprehensive dataset is needed. For that, a real-world setup is needed. To collect data, different jammers should be considered. Data collection should be done under both scenarios, under jamming and under the normal scenario. The built dataset can be used to train and test deep learning techniques. Then, these techniques can be combined with sensing to detect the strategy of the jammer and actively select the communication channel that is not under jamming attacks. Deep learning models have to be trained to recognize a jamming signal from a legitimate user on the fly. Only if that can be performed, then the legitimate user can identify the pattern of the jammer and dedicate its strategy and accordingly define a mitigation hopping.

## VI. CONCLUSION

In this paper, we presented an in-depth study of the vulnerability of 5G wireless systems to smart jamming attacks. We reviewed different types of jammers, and we showed that 5G NR enhanced the resilience of wireless cellular network to jamming attacks, primarily because of its flexible and dynamic resource allocation, yet 5G NR is still far from being secure against jamming attacks. The anti-jamming strategies presented in this paper address simple jamming attacks and are not very suitable for 5G NR. Thus, further research studies on anti-jamming techniques are highly needed.